\documentstyle[11pt]{article}
\normalbaselines
\begin{document}
\bibliographystyle{unsrt}

\def\d {\delta}
\def\l {\lambda}
\vbox{\vspace{38mm}}

Unitary symmetry and the associated Lie Algebra has played a crucial
role in atomic, nuclear and particle physics. Particularly in the
context of Grand Unification in the theory of elementary particles
and their interactions one often confronts the necessity of using
groups of higher order. It is, therefore, useful to have a general
algorithm for computing various quantities, for instance,
traces of products of generators which arise in the calculation of
higher order Feynman diagrams involving loops. Customarily one uses
the more familiar complex hermitian Gell-Mann matrices \cite{gel62} but in
many situations it is more convenient to work with real Okubo type
\cite{oku62} generators.

The generators of $SU(N)$ in the fundamental representation may be
chosen to be
\begin{equation}
(A^I_J)^K_L=\d ^I_L\d ^K_J-{1\over N}\d ^I_J \d ^K_L
\end{equation}
where $A^I_J$ are real but not all hermitian; the indices $I$, $J$
which label the generators run from 1 to $N$. The indices $K$ and
$L$, also running from 1 to $N$, label the elements of the matrix
representation of the generators. Since only $N^2-1$ of them are
independent, the Cartan matrices satisfy automatically the constraint
condition
\begin{equation}
A^I_I=0
\end{equation}
where a sum over repeated indices is implied here and elsewhere in this
note unless explicitly stated otherwise. Also all the $A$'s are traceless
as should be the case for the generators of a unimodular group. The
structure constants $(F^P_Q)^{KI}_{JL}$ obtained from
\begin{equation}
[A^I_J,A^K_L]^M_N=(F^P_Q)^{KI}_{JL}(A^Q_P)^M_N,
\end{equation}
are real and expressible as
\begin{equation}
(F^P_Q)^{KI}_{JL}=\d ^P_J\d ^K_Q\d ^I_L-\d ^P_L\d ^I_Q\d ^K_J,
\end{equation}
and are antisymmetric under the simultaneous exchange of the upper and
lower pairs of indices
\begin{equation}
(F^P_Q)^{KI}_{JL}=-(F^P_Q)^{IK}_{LJ}.
\end{equation}
For $P=Q$ obviously $F=0$ and
\begin{equation}
(F^P_Q)^{JI}_{JL}=\d ^P_Q\d ^I_L-N\d ^P_L\d ^I_Q,
\end{equation}
and a similar relation follows if $I$ is put equal to $L$. Of course these
$F$-matrices satisfies the Jacobi identity
\begin{equation}
(F^P_Q)^{MK}_{LN}(F^W_X)^{QI}_{JP}+(F^P_Q)^{IM}_{NJ}(F^W_X)^{QK}_{LP}+
(F^P_Q)^{KI}_{JL}(F^W_X)^{QM}_{NP}=0,
\end{equation}
as can be directly verified with the help of eq.(4). Working out the
anticommutation relations for the $A$'s requires a little more effort
and leads to
\begin{equation}
\{A^I_J,A^K_L\} ^M_N=(D^P_Q)^{KI}_{JL}(A^Q_P)^M_N+{2\over N} (A^I_J)
^K_L\d ^M_N,
\end{equation}
where
\begin{equation}
(D^P_Q)^{KI}_{JL}=\d ^I_L\d ^K_Q\d ^P_J+\d ^K_J\d ^I_Q\d ^P_L -
{2\over N}(\d ^I_J\d ^K_Q\d ^P_L+\d ^I_Q\d ^P_J\d ^K_L-\d ^I_L
\d ^K_J\d ^P_Q),
\end{equation}
which may be observed to vanish identically for $I=L$ and $J=K$ leaving
\begin{equation}
\{ A^I_J,A^J_I\}={2\over N}(N^2-1){\bf 1}.
\end{equation}
The trace of the product of an arbitrary number of $A$-matrices is not
difficult to deduce. Calling Kr\"onecker deltas involving upper and
lower indices as proper combinations we can deduce that
$$Tr(A^{a_1}_{b_1}A^{a_2}_{b_2}\cdots A^{a_n}_{b_n})=\d ^{a_1}_{b_2}
\d ^{a_2}_{b_3}\cdots \d ^{a_n}_{b_1}$$
$$-{1\over N}\Big [ \d ^{a_1}_{b_1}\d ^{a_2}_{b_3}\d ^{a_3}_{b_4}\cdots
\d ^{a_n}_{b_2}+\Big\{ \pmatrix{n\cr 1\cr}-1\Big\}\ {\rm proper\
combinations}\Big ]$$
$$+{1\over N^2}\Big [ \d ^{a_1}_{b_1}\d ^{a_2}_{b_2}\d ^{a_3}_{b_4}\cdots
\d ^{a_n}_{b_3}+\Big\{ \pmatrix{n\cr 2\cr}-1\Big\}\ {\rm proper\
combinations}\Big ]$$
\begin{equation}
+\cdots (-1)^{n-1}(n-1){1\over N^{n-1}}(\d ^{a_1}_{b_1}\d ^{a_2}_{b_2}
\cdots \d ^{a_n}_{b_n}).
\end{equation}
It may be noted that
(A^I_J)^P_Q(A^J_I)^R_S=\d ^R_Q\d ^P_S-{1\over N}\d ^P_Q\d ^R_S
=(A^P_Q)^R_S
\end{equation}
and that the Casimir operators of the group are easily written down
in this schema through complete contractions
\begin{equation}
(A,A)\equiv A^I_J A^J_I
\end{equation}
\begin{equation}
(A,A,A)\equiv A^I_J A^J_K A^K_I
\end{equation}
and so on. For $SU(N)$ contractions of products of $A$'s greater than
$N$ in number give scalars that are not independent of the preceeding
constructions.

It is also sometimes useful to link the generators exhibited in the Okubo
form in terms of `annihilation' and `creation' operators $a_J$ and $a_I
^{\dag}\equiv a^I$ \cite{lip65} which satisfy the commutation relations
\begin{equation}
[a_I,a^J]=\d ^J_I\ \ [a_I,a_J]=0=[a^I,a^J]
\end{equation}
by observing that
\begin{equation}
A^I_J\equiv a^I a_J-{1\over N}\d ^I_J a^S a_S
\end{equation}
satisfy all the properties required of the generators of $SU(N)$. It may
be remarked that $a_J$ and $a_I^{\dag}$ could also have been taken to
obey anti-commutation relations without affecting the commutation relation
of the bilinears.

We now proceed to illustrate through some examples derived from the
$SU(3)_c$ sector of quantum chromodynamics the efficacy of the adopted
representation. First, let us write down some relevant equations for
$SU(3)$, which are only special cases of the equations written before.
The Okubo $A$'s and the Gell-Mann $\lambda$'s are related by
\begin{eqnarray}
A^1_2={1\over 2}(\l _1-i\l _2)& A^2_1={1\over 2}(\l_1+i\l_2)& A^1_3=
{1\over 2}(\l_4-i\l_5)\nonumber\\
A^3_1={1\over 2}(\l_4+i\l_5)& A^2_3={1\over 2}(\l_6-i\l_7)& A^3_2=
{1\over 2}(\l_6+i\l_7)\nonumber\\
A^1_1={1\over 2}(\l_3+{1\over \sqrt{3}}\l_8)& A^2_2={1\over 2}(-\l_3
+{1\over \sqrt{3}}\l_8)& A^3_3=-{1\over\sqrt{3}}\l_8
\end{eqnarray}
and thus it is worth remarking that $A^I_J$'s ($I\not= J$) are the
raising and lowering operators for the $I$, $U$ and $V$-spins
corresponding to the three $SU(2)$ subgroups. It may also be observed
that under transposition $(A^I_J)^T=(A^J_I)$. The following properties
may readily be deduced:
\begin{equation}
Tr(A^I_J)=0
\end{equation}
\begin{equation}
Tr(A^I_JA^J_I)=8
\end{equation}
\begin{equation}
Tr(A^I_J A^K_L)=(A^I_J)^K_L=(A^K_L)^I_J
\end{equation}
\begin{equation}
Tr(A^I_J A^K_I A^J_K)={56\over 3}
\end{equation}
\begin{equation}
Tr(A^I_J A^J_I A^K_L A^L_K)={64\over 3}
\end{equation}
\begin{equation}
Tr(A^I_J A^J_K A^K_L A^L_I)={32\over 3}.
\end{equation}
In QCD, the quark-gluon interaction term in the Lagrangian density is
usually depicted as
\begin{equation}
\cal{L} _{q\bar q g}=-ig_3\bar q\gamma_{\mu}{\l ^a\over 2}qG_a^{\mu}
\end{equation}
which we shall write in the Okubo scheme as
\begin{equation}
\cal{L} _{q\bar q g}=-ig'_3{\bar q}^P\gamma_{\mu}(A^I_J)^Q_Pq_Q
G^{J\mu}_I
\end{equation}
where $P$ and $Q$ play the role of color labels of the quarks. A comparison
of eqs.(22) and (23) yields
\begin{equation}
g'_3={1\over 2}g_3.
\end{equation}
Also the contraction of a pair of Okubo gluon fields yields a factor of two
as compared to what one obtains with the usual definition of the fields. To
expose some of the advantages of the present scheme let us augment the
Standard Model by introducing a set of color-octet scalar bosons $\chi^I
_J$. Such an extension is indicated as a possibility \cite{dai92, kun93}
if one wishes to break the electroweak symmetry dynamically through the
onset of a $\bar t t$ condensate while keeping the top mass within the
experimental bound in the context of a composite scale of the order of
$M_{GUT}\sim 10^{15}$ GeV. The Lagrangian density describing the
interaction of the color-octet scalar with the quarks is
\begin{equation}
\cal{L} _{\chi q \bar q}=-ig'_t\bar\psi^{iP}_L(A^I_J)^Q_P(t_R)_Q\chi^J
_{iI}\ +\ {\rm h.c.},
\end{equation}
where $i$ is the $SU(2)_L$ index, $I$, $J$, $K$, $L$ are the $SU(3)$
indices, the letters $R$ and $L$ stand for right and left and $g'_t$
is the relevant Yukawa coupling. In the minimal model considered above
we have taken the $\chi$ to interact only with the third generation of
quarks, its effect on the other quarks being through the
Cabibbo-Kobayashi-Maskawa mixing.

Now if we were to calculate the contribution of the quark box diagram
to the $\chi\chi\rightarrow\chi\chi$ scattering, it would in the
conventional approach have involved tedious manipulations involving
$d$ and $f$ coefficients while in the Okubo format the relevant term in
the fourth order of the perturbation expansion being of the form
$$
(\bar\psi_L^{i_1P_1}(A^{I_1}_{J_1})^{Q_1}_{P_1}(t_R)_{Q_1}\chi^{J_1}
_{i_1I_1}).(\bar t_R^{Q_2}(A^{I_2}_{J_2})^{P_2}_{Q_2}(\psi_{Li_2})
_{P_2}\chi^{i_2J_2}_{I_2}).
$$
\begin{equation}
(\bar\psi_L^{i_3P_3}(A^{I_3}_{J_3})^{Q_3}_{P_3}(t_R)_{Q_3}\chi^{J_3}
_{i_3I_3}).(\bar t_R^{Q_4}(A^{I_4}_{J_4})^{P_4}_{Q_4}(\psi_{Li_4})
_{P_4}\chi^{i_4J_4}_{I_4})
\end{equation}
the spinorial contractions (occurring in the box diagram contribution)
yield merely terms involving products of Kr\"onecker deltas such as
for example
\begin{equation}
\d^{i_1}_{i_2}\d^{i_3}_{i_4}\d^{P_1}_{P_2}\d^{P_3}_{P_4}
\d^{Q_2}_{Q_3}\d^{Q_4}_{Q_1}
\end{equation}
leading typically to expressions of the form
\begin{equation}
(A^{I_1}_{J_1})^{Q_1}_{P_1} (A^{J_2}_{I_2})^{P_2}_{Q_2}
(A^{I_3}_{J_3})^{Q_3}_{P_3} (A^{J_4}_{I_4})^{P_4}_{Q_4}
\chi^{J_1}_{i_1I_1}\chi^{i_1J_2}_{I_2}
\chi^{J_3}_{i_3I_3}\chi^{i_3J_4}_{I_4}
\end{equation}
which by the grace of Okubo immediately yields
\begin{equation}
\chi^{J_1}_{i_1J_4}\chi^{i_1J_2}_{J_1}
\chi^{J_3}_{i_3J_2}\chi^{i_3J_4}_{J_3}
\end{equation}
as   $\chi^I_I=0$.   In  an  analogous  manner  the  amplitude  for
$gg\rightarrow gg$ may also be evaluated. As another example  the
contribution  of  the colored fields to the renormalization group
equation for the $\bar t t \chi$ vertex yields for  the  one-loop
$\beta$-function
\begin{equation}
16\pi^2{dg'_t\over   dt}={3\over  4}g_t^2g'_t+{2\over  3}{g'_t}^3
-{16\over 3}g_3^2g'_t
\end{equation}
where the factors occuring in  the  different  terms  are  merely
results  of  simple  manipulations with Kr\"onecker deltas rather
than  relatively  more  complicated  gymnastics   involving   the
products of three $\l$ matrices and the $d$ and $f$ coefficients.
Thus  it would appear that the Okubo choice of representation for
the generators of unitary unimodular groups  provides in many
situations a  simpler
scheme   requiring   manipulations   involving  objects  no  more
complicated than Kr\"onecker deltas  and,  therefore,  affords  a
distinct advantage in computations.

\end{document}